\documentclass[10pt]{bmc_article}

\usepackage{cite} 
\usepackage{url}  
\usepackage{ifthen}  
\usepackage{multicol}   
\usepackage[utf8]{inputenc} 
\def\includegraphics{}

\usepackage{graphicx}

\newboolean{publ}

\newenvironment{bmcformat}{\begin{raggedright}\baselineskip20pt\sloppy\setboolean{publ}{false}}{\end{raggedright}\baselineskip20pt\sloppy}

\begin{document}
\begin{bmcformat}


\title{An Outbreak Vector-host Epidemic Model with Spatial Structure: The 2015 Zika Outbreak in Rio De Janeiro}
 

\author{W.E. Fitzgibbon$^{1}$%
       \email{W.E. Fitzgibbon - wfitzgib@Central.UH.EDU}%
      \and
         J.J. Morgan$^2$%
         \email{J.J. Morgan - jjmorgan@Central.UH.EDU}
       and 
         G.F. Webb\correspondingauthor$^3$%
         \email{G.F. Webb - glenn.f.webb@vanderbilt.edu}%
      }


\address{%
   \iid(1)Department of Mathematics, University of Houston,%
        77204 Houston, TX, USA\\
    \iid(2)Department of Mathematics, University of Houston,%
        77204 Houston, TX, USA\\
    \iid(3)Department of Mathematics, Vanderbilt University, %
        37240 Nashville, TN, USA
}%

\maketitle


\begin{abstract}
        \paragraph*{Background:} A deterministic model is developed for the spatial spread of an epidemic disease in a geographical setting. The disease is borne by vectors to susceptible hosts through criss-cross dynamics. The model is focused on an epidemic outbreak that initiates from a small number of cases in a small sub-region of the geographical setting. 
      
       \paragraph*{Methods:} Partial differential equations are formulated to describe the interaction of the model compartments.
        
         \paragraph*{Results:} The partial differential equations of the model are analyzed and proven to be well-posed. The epidemic outcomes of the model are correlated to the  spatially dependent parameters and initial conditions of the model.  
         
           \paragraph*{Conclusions:} A version of the model is applied to the 2015-2016 Zika outbreak in the Rio de Janeiro Municipality in Brazil.
\end{abstract}

\ifthenelse{\boolean{publ}}{\begin{multicols}{2}}{}


\section{Background}
The model describes an outbreak epidemic with host-vector population dynamics in a geographical region. The epidemic outbreak begins at time 0 in a small sub-region in which the epidemic disease is not yet present. The goal of the model is to aid understanding of how the introduction of a very small number of cases in a specific location in the geographic region will result in a dissipation or a sustained and growing epidemic. The focus of the study is on the importance of spatial effects in these possible outcomes. If the equations of the model do not depend on spatial considerations, then a corresponding system of ordinary differential equations can be analyzed for their asymptotic behavior (Appendix).

The geographical region is denoted by $\Omega \subset R^2$. The background population of uninfected hosts in 
$\Omega$ has geographic density $H_u(x,y)$, which is unchanging in time in the demographic and epidemic context of the outbreak. Thus, the model is viewed as applicable to an early phase of the epidemic, during which the epidemic does not alter the basic geographic and demographic population structure of hosts, and the susceptible host population is not altered significantly by immunity to re-infection.

\subsection{Compartments of the Model}
The model consists of the following compartments:  
\vspace{0.1in}
\begin{itemize}
\item[ ] The density of infected hosts $H_i(t,x,y)$ at time $t$ at $(x,y) \in \Omega$, with initial condition $H_{i0}(x,y)$.
\item[ ] The density of uninfected vectors $V_u(t,x,y)$ at time $t$ at $(x,y) \in \Omega$, with initial condition 
$V_{u0}(x,y)$.
\item[ ] The density of infected vectors $V_i(t,x,y)$ at time  $t$ at $(x,y) \in \Omega$, with initial condition $V_{i0}(x,y)$.
\end{itemize}

The initial state $H_{i0}(x,y)$ consists of a relatively small number of infected hosts  located at time $0$ in a small sub-region of $\Omega$.
This input corresponds to an arrival of  infected hosts from outside $\Omega$. This input is assumed to have a threshold level,  which may include multiple cases produced from arriving cases. The background uninfected mosquito population has an initial state $V_{u0}(x,y)$, which decreases as the infected vector population increases. The initial population of infected vectors $V_{i0}(x,y)$ in $\Omega$ is assumed proportional to $H_{i0}(x,y)$. 

\subsection{Equations of the Model}
The equations of the model in the case that transmission from vectors to hosts is year-round are 
\begin{eqnarray}
\label{ZIKAmodel}
\frac{\partial}{\partial t} H_i(t,x,y)  &=&  \nabla \cdot \delta_1(x,y)  \nabla H_i(t,x,y) 
  - \lambda(x,y) \, H_i(t,x,y) \label{Eq-1} \\  
  &+&   \sigma_1(x,y) \, H_u(x,y) \,  V_i(t,x,y),   \nonumber \\
\frac{\partial}{\partial t} V_u (t,x,y)  &=& \nabla \cdot \delta_2(x,y)  \nabla V_u(t,x,y)
  - \sigma_2(x,y) V_u(t,x,y) H_i(t,x,y) \label{Eq-2} \\
&+& \beta(x,y) \bigg( V_u(t,x,y) + V_i(t,x,y) \bigg)  \nonumber \\ 
&-& \mu(x,y) \bigg( V_u(t,x,y) + V_i(t,x,y) \bigg) V_u(t,x,y), \nonumber \\
 \frac{\partial}{\partial t} V_i (t,x,y)  &=&   \nabla \cdot \delta_2(x,y)  \nabla V_i(t,x,y) 
+ \sigma_2(x,y) V_u(t,x,y) H_i(t,x,y)   \label{Eq-3}  \\
& &  \quad \quad - \mu(x,y) \bigg( V_u(t,x,y) + V_i(t,x,y) \bigg) V_i(t,x,y). \nonumber
\end{eqnarray}

\subsection{Well-posedness of the Model}
{\bf Theorem.}
Let $\Omega$ be a bounded domain in $R^2$ with smooth boundary $\partial \Omega$ such that $\Omega$ lies locally on one side of $\partial\Omega$.
Let $\beta$, $\mu$,  $\lambda$, $\sigma_1$, $\sigma_2, \delta_1, \delta_2 
 \, \in C_+^0(\overline{\Omega})$,  and let 
$H_{u}, I_0, V_{u0} ,V_{i0} \in C_+^1(\overline{\Omega})$.
There exists a unique global classical solution $\{H_i(t),V_u(t),V_i(t)\} \in C_{+}^{1}(\overline{\Omega}), \, t \geq 0$, to (\ref{Eq-1}),(\ref{Eq-2}),(\ref{Eq-3}), satisfying  boundary  conditions 
$$
\frac{\partial}{\partial \eta} H_i(t,x,y) = 0, \,
\frac{\partial}{\partial \eta} V_u(t,x,y) = 0, \,
\frac{\partial}{\partial \eta} V_i(t,x,y) = 0, \, (x,y) \in \partial \Omega, \, t >0
$$
and initial conditions
$$
H_i(0,x,y) = H_{i0}(x,y), \, V_u(0,x,y) = V_{u0}(x,y), \,  V_i(0,x,y) = V_{i0}(x,y), \, (x,y) \in  \Omega.
$$

{\bf Proof.} 
We first observe that a unique classical solution $\{H_i(t),V_u(t),V_i(t)\}$ exists in $C^{1}(\overline{\Omega})$
on a maximal interval of existence $[0,T_{max})$ (\cite{Martin}, \cite{Pazy}, \cite{Smoller}).
Standard arguments (\cite{Smoller}) guarantee that $\{H_i(t),V_u(t),V_i(t)\}$ remain  nonnegative for $t \in [0,T_{max})$.
Moreover, the classical solution can be globally defined if we can establish uniform \textit{a priori} bounds. 
Set $M(t,x,y) = V_u(t,x,y) + V_i(t,x,y)$ and add equations (\ref{Eq-2}) and (\ref{Eq-3}) to obtain
\begin{eqnarray}
\frac{\partial}{\partial t} M(t,x,y)   
&=& \nabla \cdot \delta_2(x,y)  \nabla M(t,x,y)  \label{Eq-4} \\
&+& \, \,  \beta(x,y) M(t,x,y)\, \,  - \, \, \mu(x,y) M(t,x,y)^2. \nonumber
\end{eqnarray}
Theorem 1 in \cite{Fitz-Langlais} guarantees the existence of a unique global classical solution 
$M(t) \in C_+^1(\overline{\Omega})$ to Equation  (\ref{Eq-4}) satisfying
$$
\frac{\partial}{\partial \eta} M(t,x,y) = 0, \, (x,y) \in \partial \Omega,  \, t \geq 0, \, M(0,x,y) = V_{u0}(x,y) + V_{i0}(x,y), \, (x,y) \in  \Omega.
$$
Further, in \cite{Fitz-Langlais} it is proved that there exists $\overline{M} \in C_+^0(\overline{\Omega})$, $\overline{M} \neq 0$, such that $\lim_{t \rightarrow \infty}M(t) = \overline{M} \in C_+^0(\overline{\Omega})$.  We note that the disease free equilibrium of (\ref{Eq-1}),(\ref{Eq-2}),(\ref{Eq-3}) is $(0,\overline{M},0)$. From \cite{Fitz-Langlais} there exists $N_1 > 0$ such that 
$max_{t \geq 0} \| M(t) \|_{C_+^0(\overline{\Omega})} < N_1$, which implies  $\| V_i(t) \|_{C_+^0(\overline{\Omega})}, \,
\| V_u(t) \|_{C_+^0(\overline{\Omega})} < N_1$. Then, since $\lambda > 0$ in (\ref{Eq-1}), there exists $N_2 > 0$ such that $\| H_i(t) \|_{C_+^0(\overline{\Omega})} < N_2$. Consequently, the solution exists globally on $[0,\infty)$.

\section{The Spatially Structured Basic Reproduction Number}
Define the basic reproduction number of the model (\ref{Eq-1}),(\ref{Eq-2}),(\ref{Eq-3}) as
$$R_0(x,y) =  \frac{\sigma_1(x,y) \sigma_2(x,y) H_u(x,y)}{\lambda(x,y) \mu(x,y)}.$$
$R_0(x,y)$ is interpreted as the average number of new cases generated by a single case at a given  location $(x,y)$ 
in $\Omega$. Our motivation for this definition is the basic reproduction number $R_0$ of the spatially independent model (Appendix). Simulations of the spatially dependent model show that for certain parameterizations of equations (\ref{Eq-1}),(\ref{Eq-2}),(\ref{Eq-3}), the solutions have the following behavior:  
(1) If $R_0(x,y) < 1$ everywhere in $\Omega$, then the populations of both infected hosts and infected vectors extinguish, and the populations converge to the disease free equilibrium.  
(2) If $R_0(x,y) > 1$ in some sub-region $\Omega_0 \subset \Omega$, then the populations of both infected hosts and infected vectors converge to an endemic equilibrium independently of the initial conditions, and even if the average value of $R_0(x,y)$ in all of $\Omega$ is $ < 1$.

\section{The 2015 Zika Outbreak in Rio de Janeiro Municipality}
We apply a version of the model (\ref{Eq-1}),(\ref{Eq-2}),(\ref{Eq-3}) to the 2015 Zika outbreak in Rio de Janeiro Municipality in Brazil. Because disease transmission in the Municipality is seasonal, equations (\ref{Eq-2}) and (\ref{Eq-3}) must be modified to account for seasonality. We assume that there is a mosquito population breeding term 
$\beta(t,x,y)$, depending on time. We also assume that there is an on-going mosquito loss term $\mu_1(x,y)$, corresponding to the average mosquito life-span $1 / \mu_1(x,y)$, independent of the carrying capacity loss term $\mu(x,y)$. The modified equations are

\begin{eqnarray}
\label{ZIKAmodelModified}
\frac{\partial}{\partial t} V_u (t,x,y)  &=& \nabla \cdot \delta_2(x,y)  \nabla V_u(t,x,y)
  - \sigma_2(x,y) V_u(t,x,y) H_i(t,x,y) \label{Eq-2mod} \\
&+& \beta(t,x,y) \bigg( V_u(t,x,y) + V_i(t,x,y) \bigg) - \mu_1(x,y) V_u(t,x,y)  \nonumber \\ 
& & \quad - \mu(x,y) \bigg( V_u(t,x,y) + V_i(t,x,y) \bigg) V_u(t,x,y), \nonumber \\
 \frac{\partial}{\partial t} V_i (t,x,y)  &=&   \nabla \cdot \delta_2(x,y)  \nabla V_i(t,x,y) 
+ \sigma_2(x,y) V_u(t,x,y) H_i(t,x,y)   \label{Eq-3mod}  \\
& &  - \mu(x,y) \bigg( V_u(t,x,y) + V_i(t,x,y) \bigg) V_i(t,x,y) \nonumber -  \mu_1(x,y) V_i(t,x,y). \nonumber
\end{eqnarray}

The host population are the people in the Municipality, which in 2016 is approximately 6,000,000, in a geographical region of approximately 1,200 square kilometers (Source: \textit{Instituto Brasileiro de Geografia e Estatistica}). The vector population is the female \textit{Aedes aegypti} mosquito. The Municipality comprises 33 sub-districts, with population densities ranging from 1,000 to 50,000 inhabitants per square kilometer (Figure 1). 

A small number of cases were recorded in the Municipality into the summer of 2015, with the highest number of cases in the eastern region of the Municipality (\cite{Brasil}, \cite{Honorio}).
The Brazilian Health Ministry (\cite{Boletim13},\cite{Boletim16}) reported that Rio de Janeiro State (population approximately 16,000,000) registered a count of 25,930 cumulative  cases from January 1, 2016  to April 1, 2016 (with incidence of 156.7 cases per 100,000 inhabitants),  and 32,312 cumulative cases  by April 23, 2016 (with incidence of 195.2 cases per 100,000 inhabitants). The Ministry (\cite{Boletim17},\cite{Boletim18}) reported no new cases in the State from April 24, 2016 to May 7, 2016.
In \cite{Bastos} the weekly  case data for  Rio de Janeiro Municipality is given from November 1, 2015 through April 10, 2016, during which time the reporting of cases became mandatory. The period can be viewed as the 2015-2016 seasonal mosquito transmission period of the epidemic in the Municipality.

\subsection{Parameterization of the Rio de Janeiro Model}
We simulate this case data for the Rio de Janeiro Municipality with the following parameterization: The time units are weeks. The spatial units are kilometers and $\Omega = (-25,25) \times (-12,12)$. 
The average length of the infectious period of infected people is approximately 1 to 2 weeks and we set 
$\lambda(x,y) = 1.0$ (\cite{CDC}). The average lifespan of female \textit{Aedes aegypti} mosquitoes is approximately two weeks in an urban environment  (\cite{Brady},\cite{Kucharsky},\cite{Otero}), and we set $\mu_1(x,y) =0.5$.
The total uninfected host population is $6,000,000$, with geographical density function $H_u(x,y) = 50.0 + 10^2 \, (1.0 + \sin(0.02 \pi x) \, \cos(0.03 \pi y))$ (Figure 2A), which corresponds approximately to the population density distribution  in Figure 1. 

Set $\mu_0 = 0.0001$ and the density dependent mosquito loss function  
$\mu(x,y) = \mu_0 (1.0 +100 \, gauss(20.0,30.0,x) \times gauss(0.0,30.0,y))$ (Figure 2B), which corresponds to higher levels of mosquito control in the eastern region of the Municipality, where the population density is highest.
Here $gauss(m,sd,x)$ is the probability density function in $x$ of the normal distribution function with mean $m$ and standard deviation $sd$.
Set the transmission parameters $\sigma_1(x,y) = 0.0000051$, $\sigma_2(x,y) =0.78$ (we assume that individual mosquitoes bite multiple people, people receive multiple bites, and the probability of infection of mosquitoes is much higher than the probability of infection of people).  Set  $\delta_1 = 0.2$, which is a simplified estimate of human dispersal in urban settings. Set $\delta_2 = 0.2$, which is consistent with an estimated adult mosquito dispersal of $30 - 50 \, m$ per day (\cite{Otero}).

The time dependent mosquito breeding function is $\beta(t,x,y) = 
 \beta_0 \, emg(t,\bar{\mu},\bar{\sigma},\bar{\lambda)}$, where $\beta_0 = 300.0$ and $emg$ is the shifted exponentially modified gaussian 
$$emg(t,\bar{\mu},\bar{\sigma},\bar{\lambda)}) = \frac{\bar{\lambda}}{2} Exp\bigg(\frac{\bar{\lambda}}{2}(2 \bar{\mu} + \bar{\lambda} \bar{\sigma}^2 - 2 \,  t)\bigg)
Erfc\bigg(\frac{1}{\sqrt{2}  \, \bar{\sigma}} (\bar{\lambda} \bar{\sigma}^2 + \bar{\mu} -t)\bigg)$$
Here $Erfc$ is the complementary error function. The parameters are $\bar{\mu} = -2.0$, $\bar{\sigma} = 5.0$, 
$\bar{\lambda} = 0.2$.
The graph of the seasonal mosquito breeding function $\beta$ is given is Figure 3 ($\beta$ is independent of $x$ and $y$). 

\subsection{Initialization of the Rio de Janeiro Model}

The outbreak begins at time 0 on November 1, 2015  in a northeastern location of the Municipality with high population density. The total number of infected cases at time $0$ is $10$, with spatial distribution $H_i(0,x,y) = 10.0 \,  gauss(15.0,1.0,x) \times gauss(0.0,1.0,y)$.  At time $0$ the total number of uninfected mosquitoes is $120,000$, distributed uniformly throughout the Municipality. The total number of infected mosquitoes at time $0$ is $100$, with $V_i(0,x,y) = 10.0 H_i(0,x,y)$. 

\subsection{Simulation of the Rio de Janeiro Model}
Example 1. The simulation of the model (\ref{Eq-1}), (\ref{Eq-2mod}), (\ref{Eq-3mod}) over the time period November 1, 2015 to May 21, 2016 is graphed in Figures 4, and 5. The simulation agrees qualitatively with the  weekly reported case data for Rio de Janeiro Municipality in \cite{Bastos} (Figure 4). The spatial distributions of infected people  expand from a very small number of initial cases in a small sub-region of the Municipality, and disperses throughout the eastern region of the Municipality (Figure 5). The mosquito population rises rapidly and reaches carrying capacity at approximately 14 million in earlier 2016. During much of mosquito season, the ratio of  mosquitoes to  people is approximately $2$ to $1$, which agrees with the ratio in (\cite{Manore}).

Example 2. We repeat the simulation with the only change the location of the initial infected cases. We take $H_i(0,x,y) = 10.0 \,  gauss(10.0,1.0,x) \times gauss(-5.0,1.0,y)$ (Figure 6). The infected population again expands from the initial location and disperses throughout the eastern region of the Municipality, but at approximately one-half the number of infected cases as in Example 1. The reason is that the density of susceptible people is lower in this initial location than the initial location in Example 1. 

\section{Conclusions}
The model (\ref{Eq-1}),(\ref{Eq-2}),(\ref{Eq-3}) describes criss-cross vector-host transmission  dynamics of an epidemic outbreak in a geographical region $\Omega$.  
The outbreak occurs with a small number of infected hosts in a small sub-region of the much  larger geographical region $\Omega$. The diffusion terms describe the on-going  average spatial movement of vectors and hosts in the geographical region. The focus of the model is to describe the geographical spread from an initial localized immigration into the region, in terms of the epidemiological properties of the outbreak vector-host transmission dynamics. 

\subsection{Summary of the Outcomes of the Outbreak Model}
We prove that the partial differential equations model (\ref{Eq-1}),(\ref{Eq-2}),(\ref{Eq-3}) is mathematically well-posed, and compare its properties to an analogous ordinary differential equations model in the spatially independent case (Appendix). The outcomes of the model depend on the spatially distributed basic reproduction number 
$R_0(x,y)$.
If $R_0(x,y) < 1$ everywhere in $\Omega$, then the epidemic will extinguish.
If $R_0(x,y) > 1$ in some sub-region of  $\Omega$, then the epidemic will spread and converge to an endemic equilibrium  throughout  all of $\Omega$, independently of the location of the sub-region.

\subsection{Summary  of the Model Applied to the Zika Outbreak in Rio de Janeiro}
The model  (\ref{Eq-1}),(\ref{Eq-2}),(\ref{Eq-3}) is modified to allow seasonality of the vector population, and applied to the 2015-2016 Zika outbreak in Rio de Janeiro Municipality.  A simulation of the model provides qualitative agreement with  the case data reported by the Brazilian Ministry of Health in \cite{Bastos}. The model simulation suggests that the Zika epidemic in Rio de Janeiro Municipality, will rise each season from initial locations with very small numbers of infected people, and spread throughout the Municipality. The evolution of the epidemic depends on the initial location of infected cases. In general, the evolution of the epidemic depends on the parameterization and initialization of the model, and can be limited only by reduction of the disease parameters throughout the entire region. If the number of actual infected cases is significantly higher than the number of reported cases, then the parameterization must be adjusted accordingly.

\section*{Appendix}   
The equations (\ref{Eq-1}),(\ref{Eq-2}),(\ref{Eq-3}) without spatial dependence are 
\begin{eqnarray}
\label{ZIKA-ODEmodel}
\quad \quad \frac{d}{d t} H_i(t)&=& - \lambda H_i(t)   
+ \sigma_1 \,  V_i(t) \, H_u  \label{Eq-7X}  \\
\frac{d}{d t} V_u (t) &=&   
\beta ( V_u(t) + V_i(t) ) -\sigma_2 V_u(t) H_i(t)     
- \mu (  V_u(t) + V_i(t) ) V_u(t) \label{Eq-8X} \\
\frac{d}{d t} V_i (t) &=&   
 \sigma_2 V_u(t) H_i(t) 
- \mu ( V_u(t) + V_i(t) ) V_i(t) \label{Eq-9X}
\end{eqnarray}
with initial conditions $H_i(0) = H_{i0}$, $V_u(0) = V_{u0}$, $V_i(0) = V_{i0}$.
Set $R_0 = H_u \sigma_1 \sigma_2 / \lambda \mu$. The behavior of solutions of  equations (\ref{Eq-7X}),(\ref{Eq-8X}), (\ref{Eq-9X}) can be classified as follows:

{\bf Proposition}
If $R_0 < 1$, then the only steady states of (\ref{Eq-7X}),(\ref{Eq-8X}),(\ref{Eq-9X}) in $R_+^3$ are $ss_0 = (0,0,0)$, which  is unstable in $R_+^3$, and $ss_1 = (0,\beta/\mu,0)$, which is proportional to $\beta$ and locally exponentially asymptotically stable in $R_+^3$. 
If $R_0 < 1$, $H_i(0) > 0$, and $V_i(0) = 0$, then $(H_i(t),V_u(t),V_i(t))$ converges to $(0,\overline{M},0)$.
If $R_0 > 1$, then  $ss_0$ and $ss_1$ are unstable in $R_+^3$
and there is another steady state in $R_+^3$,
$$ss_2 = \bigg( \frac{\beta (H_u \sigma_1 \sigma_2 -\lambda \mu)}{\lambda \mu \sigma_2},
\frac{\beta \lambda}{H_u \sigma_1 \sigma_2},
\frac{\beta (H_u \sigma_1 \sigma_2 -\lambda \mu)}{H_u \mu \sigma_1  \sigma_2} \bigg)$$
$$= \bigg( \frac{\beta(R_0 -1)}{\sigma_2}, 
\frac{\beta}{R_0 \mu} , 
\frac{\lambda \beta (R_0 - 1)}{ H_u \sigma_1 \sigma_2} \bigg).$$  
which is locally exponentially asymptotically stable in $R_+^3$. 

{\bf Proof.}
It can be verified that the steady states of  (\ref{Eq-7X}),(\ref{Eq-8X}),(\ref{Eq-9X}) in $R_+^3$ are $ss_0, ss_1$, and $ss_2$. The Jacobian of (\ref{Eq-7X}),(\ref{Eq-8X}),(\ref{Eq-9X}) at $ss_0$ is

$$J(0,0,0) = \left[ \begin{array}{ccc}
-\lambda & 0 & H_u \sigma_1 \\
0 & \beta & \beta \\
0 & 0 & 0 \end{array} \right]$$
with eigenvalues $\{-\lambda,\beta,0\}$, which means that $(0,0,0)$ is unstable.

Let $M(t) = V_i(t) + V_u(t)$. Equations (\ref{Eq-8X}) and (\ref{Eq-9X}) imply $M^{\prime}(t) = \beta M(t)  - \mu M(t)^2$, $\lim_{t \rightarrow \infty}M(t) = \overline{M} = \beta / \mu$.

If $H_i(0) > 0$ and $V_i(0) = 0$, then (\ref{Eq-7X}) implies $H_i^{\prime}(0) <  0$. Assume there is a smallest positive time $t^{\ast}$ such that $H_i^{\prime}(t^{\ast}) = 0$. Then (\ref{Eq-7X}) implies $H_i(t^{\ast}) = (\sigma_1 H_u /  \lambda) V_i(t^{\ast})$. If $R_0 < 1$, then (\ref{Eq-9X}) implies 
$$
V_i^{\prime}(t^{\ast}) =  (\sigma_1 \sigma_2 H_u /  \lambda) V_i(t^{\ast}) ( M(t^{\ast}) - V_i(t^{\ast}) )
- \mu V_i(t^{\ast})  M(t^{\ast}) < - (\sigma_1 \sigma_2 H_u /  \lambda) V_i(t^{\ast})^2 < 0.
$$
Then (\ref{Eq-7X}) implies
$
H_i^{\prime \prime}(t^{\ast}) = - \lambda H_i^{\prime}(t^{\ast}) + \sigma_1 H_u V_i^{\prime}(t^{\ast}) < 0,
$
which implies $H_i(t)$ is strictly decreasing at $t^{\ast}$, yielding a contradiction. Thus, $H_i(t)$ is strictly decreasing for all $t \geq 0$. Let $H_{i,\infty} = \lim_{t \rightarrow \infty } H_i(t) \geq 0$. Assume  $H_{i,\infty} >0 $.
Then  (\ref{Eq-9X}) implies $lim_{t \rightarrow \infty } V_i(t) = \lambda H_{i.\infty} / \sigma_1 H_u > 0$. Equation (\ref{Eq-8X}) then  implies 
$\lim_{t \rightarrow \infty}V_u(t) = \beta \overline{M} / (\sigma_2 H_{i,\infty} + \mu \overline{M})$. Then 
$(H_{i,\infty},\beta \overline{M} / (\sigma_2 H_{i,\infty} + \mu \overline{M}),\lambda H_{i,\infty} / (\sigma_1 H_u))$ is a steady state of 
(\ref{Eq-7X}),(\ref{Eq-8X}),(\ref{Eq-9X}). If $R_0 < 1$, then $H_{i,\infty} =0$, yielding a contradiction. Thus, $H_{i,\infty} = 0$.

The eigenvalues of the Jacobian of (\ref{Eq-7X}),(\ref{Eq-8X}),(\ref{Eq-9X}) at $ss_1$ 
$$J(0,\beta / \mu,0) = \left[ \begin{array}{ccc}
-\lambda & 0 & H_u \sigma_2 \\
- \beta \sigma_1 / \mu & - \beta &0 \\
\beta \sigma_1 / \mu & 0 & - \beta \end{array} \right]$$
\vspace{0.1in}

\noindent 
are 
$$\{- \beta,
\frac{ - \beta - \lambda - \sqrt{(\beta - \lambda)^2 + 4 R_0 \beta \lambda}}{2},
\frac{ - \beta - \lambda + \sqrt{(\beta - \lambda)^2 + 4 R_0 \beta \lambda}}{2}\}.$$
Thus, $J(0,\beta / \mu,0)$ is unstable if $R_0 > 1$ and locally exponentially asymptotically stable if $R_0 < 1$.

The Jacobian of (\ref{Eq-7X}),(\ref{Eq-8X}),(\ref{Eq-9X}) at $ss_2$ is
$$\left[ \begin{array}{ccc}
- \lambda & 0 & H_u \sigma_1 \\
- \frac{\beta \lambda}{H_u \sigma_1} & \beta(1- \frac{\lambda \mu}{H_u \sigma_1 \sigma_1}
- \frac{H_u \sigma_1 \sigma_2}{\lambda \mu}) & \beta (1 -  \frac{\lambda \mu}{H_u \sigma_1 \sigma_2}) \\
\frac{\beta \lambda}{H_u \sigma_1} & \beta(- 2 + \frac{\lambda \mu}{H_u \sigma_1 \sigma_2}  
+\frac{H_u \sigma_1 \sigma_2}{\lambda \mu}) & \beta (- 2 + \frac{\lambda \mu}{H_u \sigma_1 \sigma_2}) \end{array} \right]
$$

$$ = \left[ \begin{array}{ccc}
- \lambda & 0 & \frac{R_0 \lambda \mu}{\sigma_2} \\
- \frac{\beta \sigma_2}{R_0 \mu} & \beta(1- \frac{1}{R_0}
- R_0) & \beta (1 -  \frac{1}{R_0}) \\
\frac{\beta \sigma_2}{R_0 \mu} & \beta(- 2 + \frac{1}{R_0}  
+ R_0) & \beta (- 2 + \frac{1}{R_0}) \end{array} \right].
$$
\vspace{0.1in}

\noindent
with eigenvalues 
$$\{- \beta,
\frac{ - R_0 \beta - \lambda - \sqrt{(R_0 \beta - \lambda)^2 + 4 \beta \lambda}}{2},
\frac{ - R_0 \beta - \lambda + \sqrt{(R_0 \beta - \lambda)^2 + 4 R_0 \beta \lambda}}{2}\}.
$$
Since 
$
- (R_0 \beta + \lambda)^2 + (R_0 \beta - \lambda)^2 + 4 \beta \lambda = - 4 (R_0 -1) \beta \lambda < 0
$
if $R_0 >1$, the eigenvalues of the Jacobian at $ss_2$ are strictly negative if $R_0 >1$, which means that $ss_2$ is locally exponentially asymptotically stable if $R_0 >1$.

\section*{Authors contributions}
    All authors conceived and developed the study.  All authors read and approved the final manuscript.  
    
{}

\newpage

\begin{figure}
\centering \includegraphics[width=5.0in,height=2.5in]{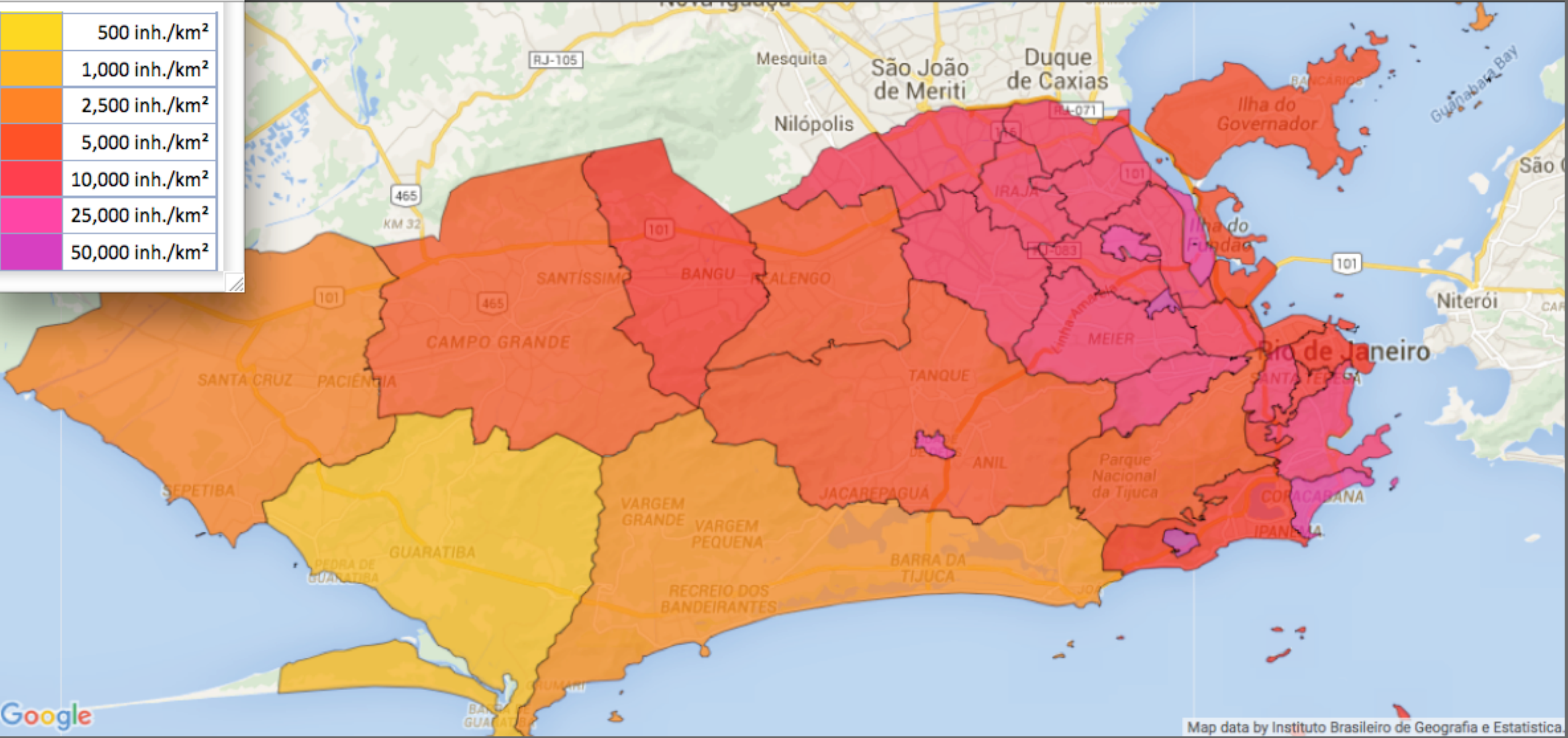}
\caption{Rio de Janeiro Municipality sub-districts. The sub-district population densities range from 1,000 to 50,000 inhabitants per square kilometer. The Municipality is approximately 50 kilometers east-west by 20 kilometers north-south, with the highest population density in the eastern region. The total population is approximately 6,000,000.  (Source: http://www.citypopulation.de/php/brazil-rio.php).} 
\end{figure}

\begin{figure}
\centering \includegraphics[width=5.6in,height=2.8in]{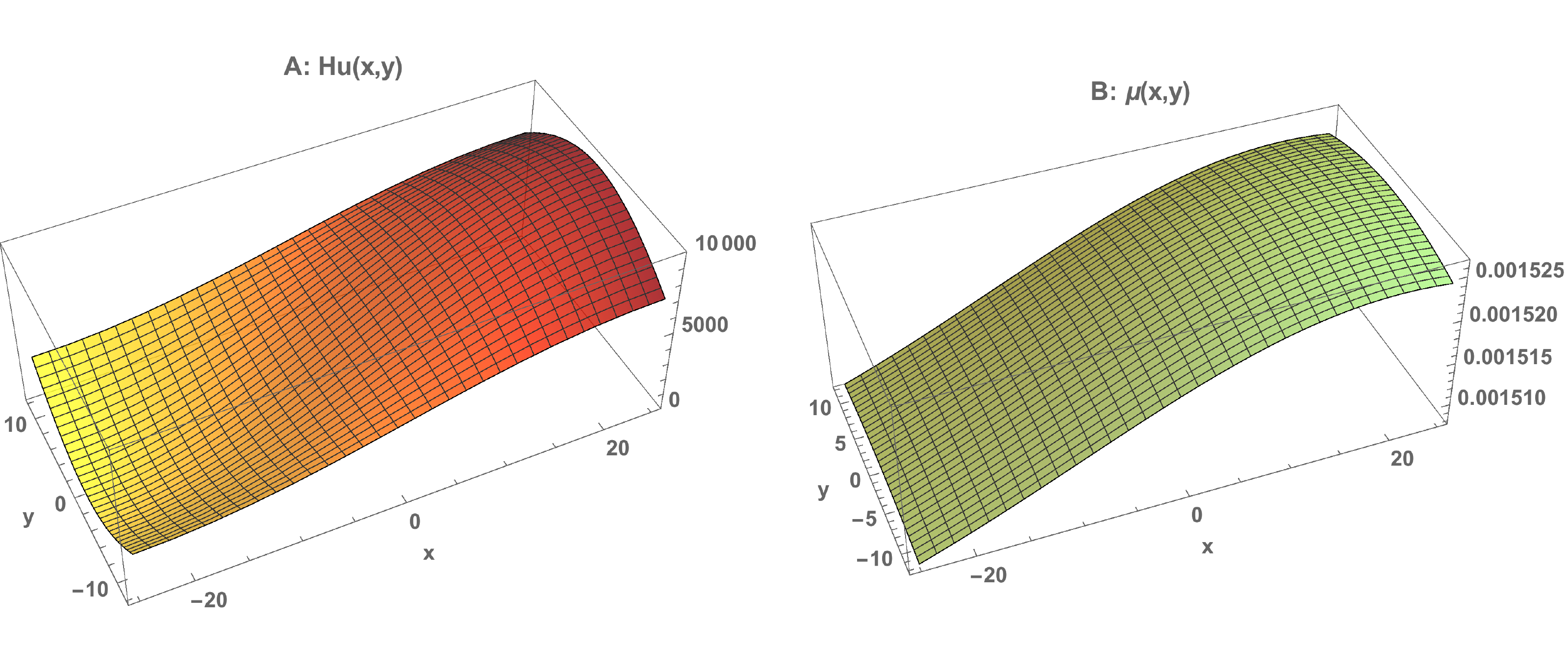}
\ \caption{A: The population of susceptible people $H_u(x,y)$ in Rio de Janeiro Municipality, which agrees approximately with the geographical population density in Figure 1.  B: The density dependent mosquito loss function $\mu(x,y)$, which is higher in locations of higher population density due to mosquito control measures.} 
\end{figure}

\begin{figure}
\centering \includegraphics[width=5.0in,height=2.4in]{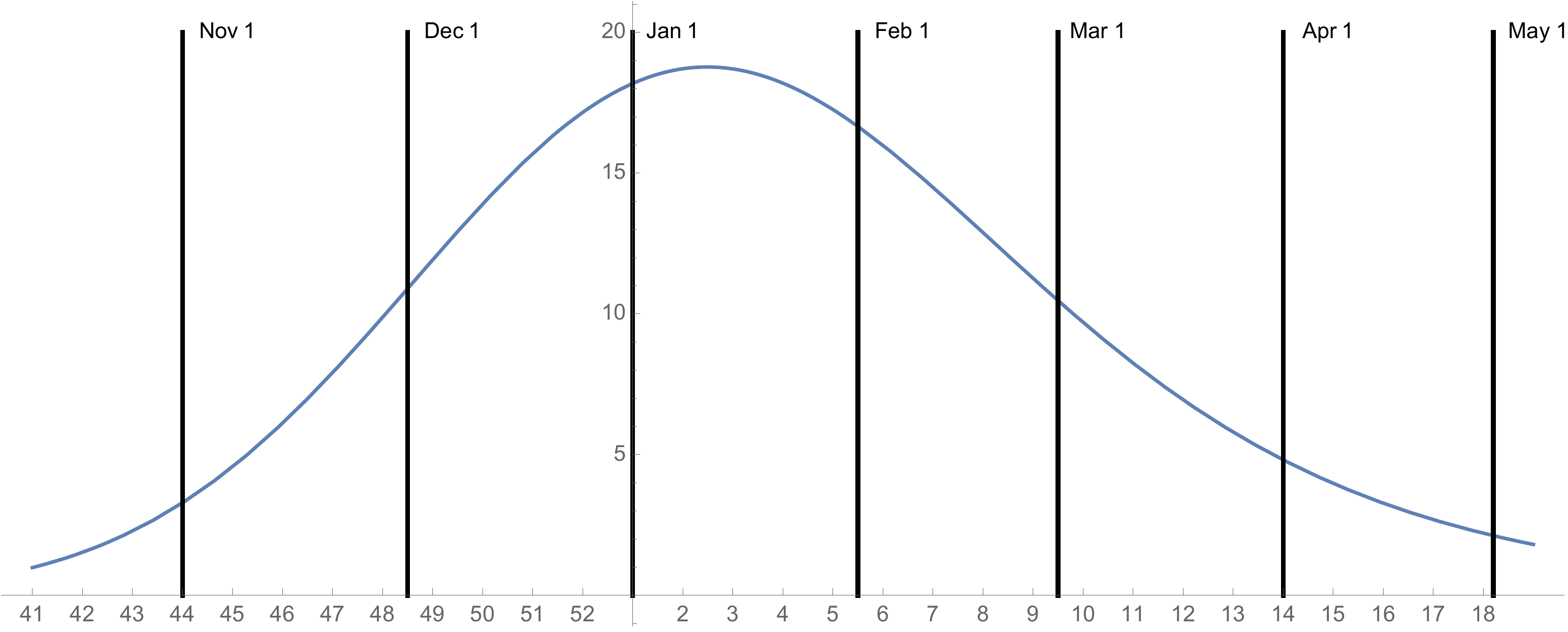}
 \caption{The time dependent mosquito breeding function $\beta(t)$ for the 2015-2016 seasonal mosquito population in Rio de Janeiro Municipality. The graph of $\beta(t)$ rises rapidly in November 2015, to its maximum in early January 2016, and then falls steadily to a low value in May 2016.}   
\end{figure}

\begin{figure}
\centering \includegraphics[width=5.0in,height=2.6in]{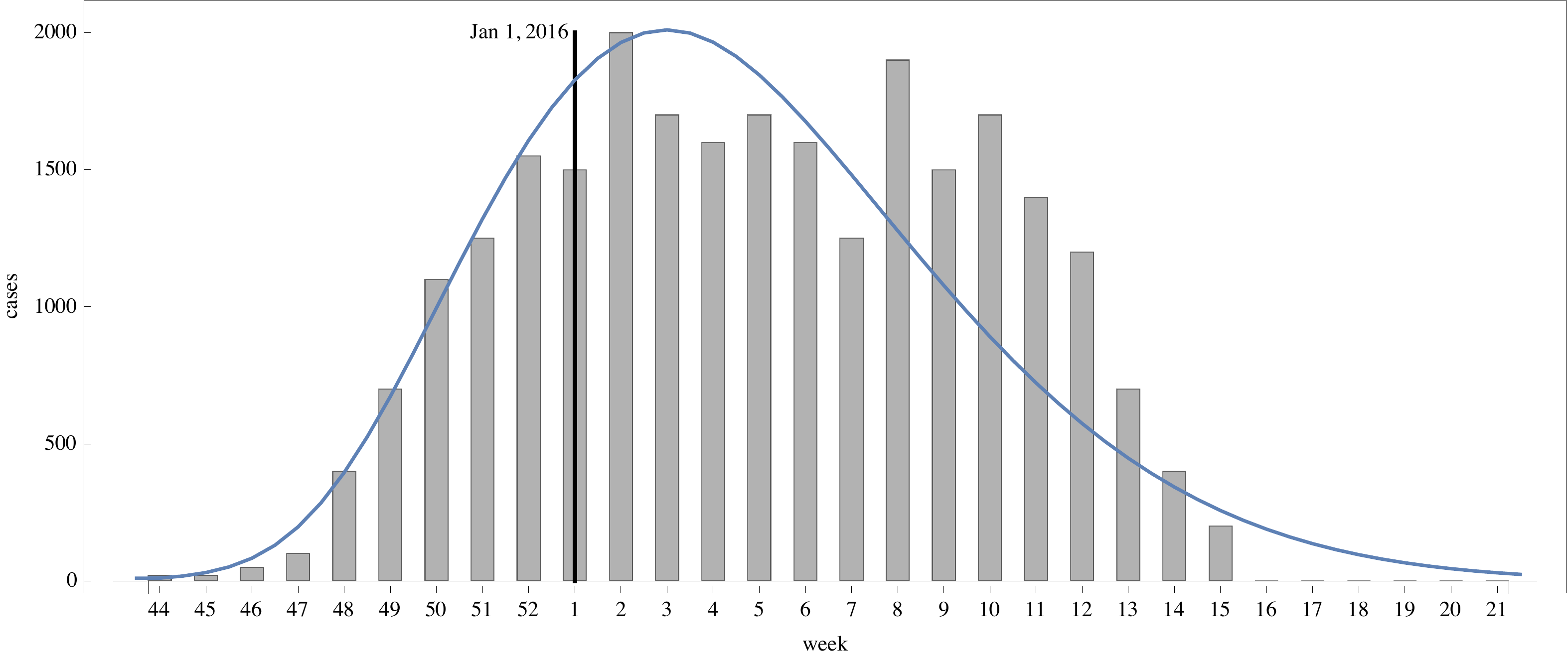}
\caption{Example 1. Simulation of the reported infected cases in the Rio de Janeiro Municipality from the beginning of the epidemic season at week 44 in  2015 to week 21 in  2016 (blue graph). The reported case values of the simulation agree qualitatively with the number of reported cases of the Brazilian Health Ministry during this period (grey bars) \cite{Bastos}.}\end{figure}

\begin{figure}
\centering \includegraphics[width=6.0in,height=2.8in]{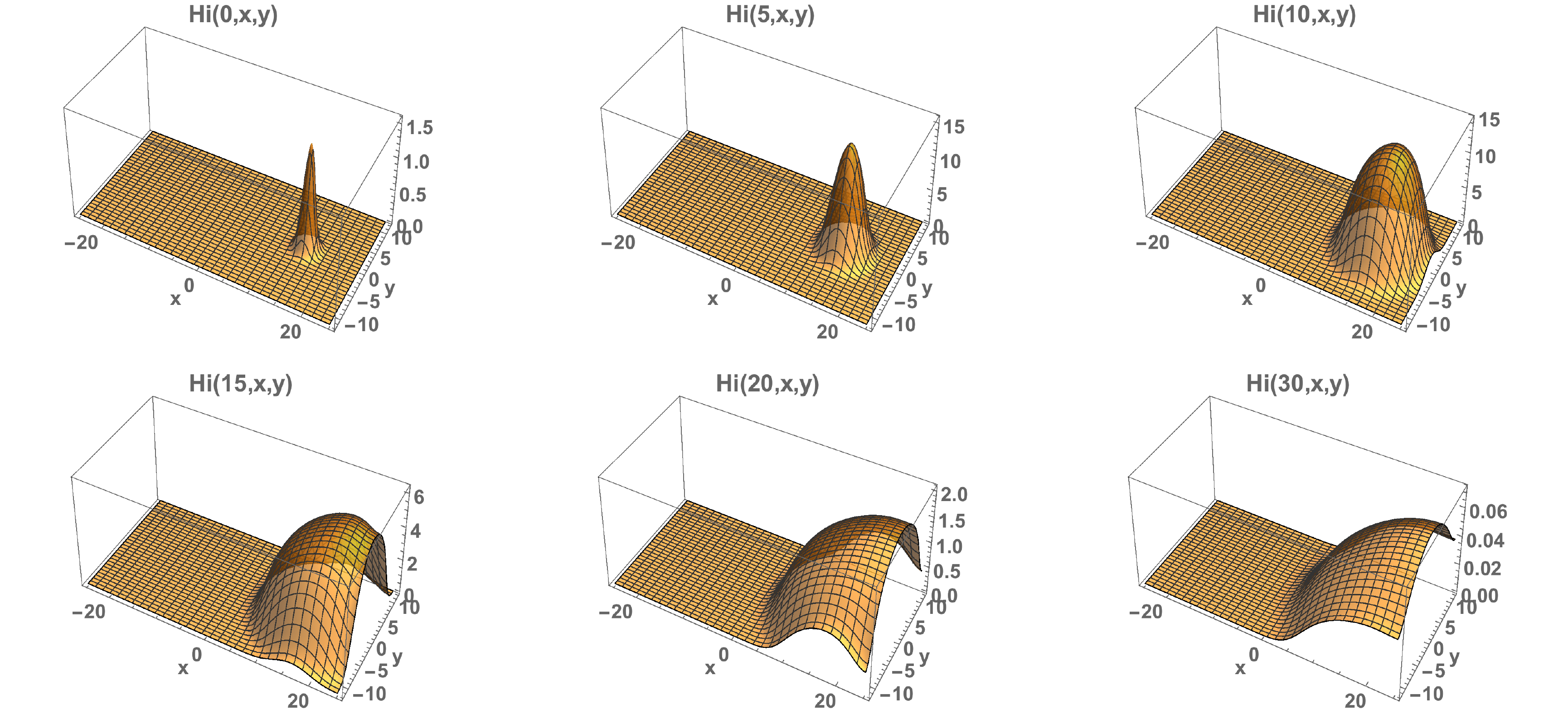}
  \caption{Example 1. Model simulation of  the spatial distribution of infected cases in the Rio de Janeiro Municipality during the 2015-2016 epidemic season. At time 0 (November 1, 2015) a very small number of  cases are located in a small region in the eastern region of the Municipality. The spatial distributions are graphed at week 5 (December 5, 2015), week 10 (January 10, 2016), week 15 (February 14, 2016), week 20 (March 21, 2016), and week 30 (April 25, 2016) from time 0. The cases concentrate in the eastern most region of the Municipality.}
\end{figure}

\begin{figure}
\centering \includegraphics[width=6.0in,height=2.8in]{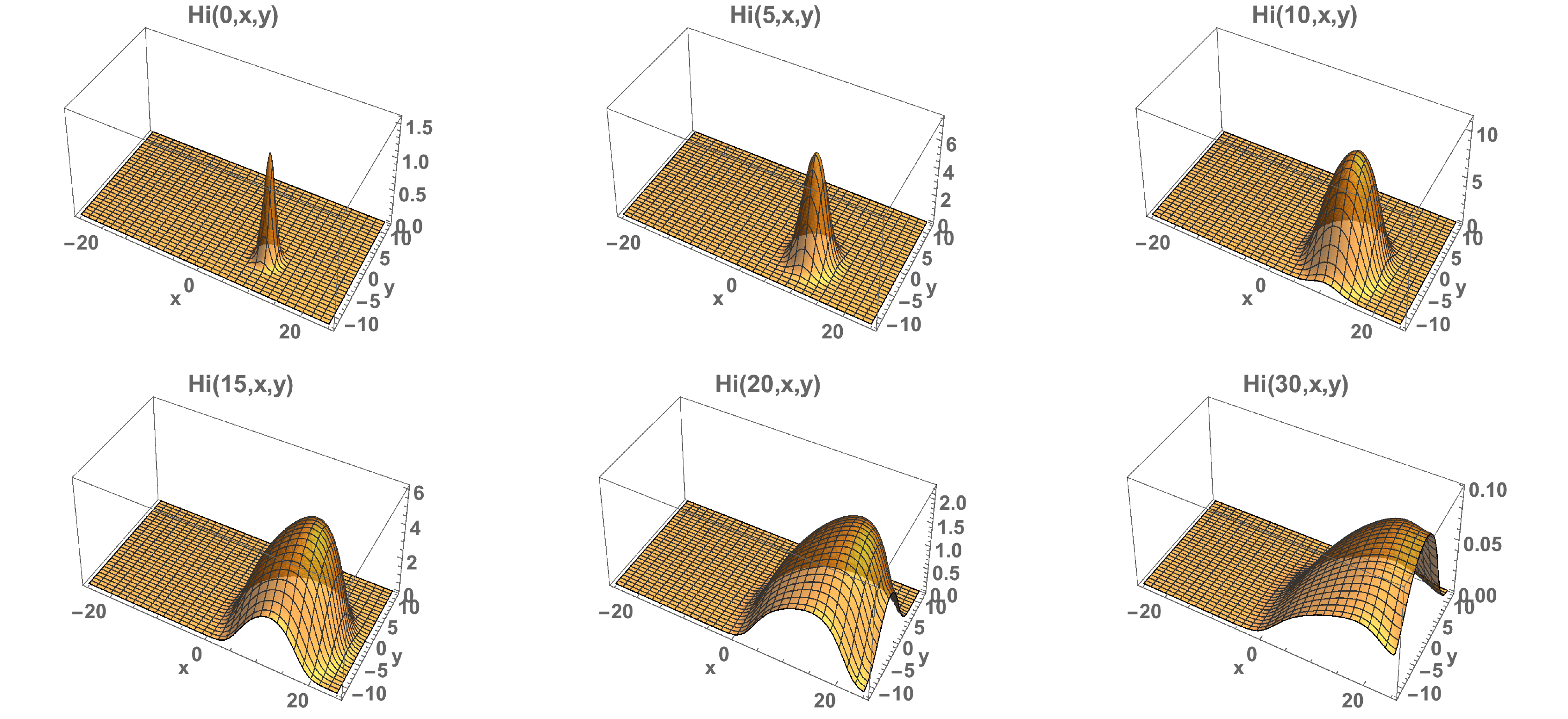}
  \caption{Example 2. Model simulation  with a change in the location of the initial infected cases. The spatial distributions are graphed at week 5 (December 5, 2015), week 10 (January 10, 2016), week 15 (February 14, 2016), week 20 (March 21, 2016), and week 30 (April 25, 2016) from time 0. The cases concentrate in the eastern most region of the Municipality, but at approximately half the number as in Example 1.}
\end{figure}

\end{bmcformat}

\end{document}